\documentclass{epl}
\usepackage{epsfig}

\begin{document}

\renewcommand{\cal}{\ensuremath{\mathcal}}
\newcommand \q  {\ensuremath{q_{EA}}}
\newcommand \be  {\begin{equation}}
\newcommand \bea {\begin{eqnarray} \nonumber }
\newcommand \ee  {\end{equation}}
\newcommand \eea {\end{eqnarray}}

\title{Crossover from stationary to aging regime in glassy
dynamics}
\author{A. Andreanov\inst{1} \and A. Lef\`evre\inst{1}}
\institute{
\inst{1}{Service de Physique Th{\'e}orique, Orme
des Merisiers -- CEA Saclay, 91191 Gif-sur-Yvette Cedex, France.}
}

\maketitle

\begin{abstract}
We study the non-equilibrium dynamics of the spherical p-spin models in
the scaling regime near the plateau and derive the corresponding
scaling functions for the correlators. Our main result is that
the matching between different time regimes fixes the
aging function in the
aging regime to $h(t)=\exp(t^{1-\mu})$. The
exponent $\mu$ is related to the one giving the length of the
plateau. Interestingly $1-\mu$ is quickly very small when one
goes away from the dynamic transition temperature in the glassy
phase. This gives new light on the interpretation of
experiments and simulations where simple aging was found to be a
reasonable but not perfect approximation, which could be attributed
to the existence of a small but non-zero stretching exponent.
\end{abstract}

\section{Introduction}
Glassy phases are widely observed in Nature and in
materials~\cite{Deb01}. The most
stricking features are very slow relaxations and history dependent
phenomena, such as rejuvenation or memory~\cite{Kov63,Joh98}. Such phenomena
reflect that
the system is aging: the older the system, the slower the
relaxation. The arrest of the dynamics is generally attributed to large scale
freezing, and the approach to the glass transition has
recently been shown to be associated with strong dynamic
heterogeneity~\cite{Hur95}.
Some regions of space may relax very fast, while others may
relax very slowly. In order to relax almost frozen regions, many
particles have to move cooperatively, and the dynamical arrest arises
when the time for relaxing large regions - the relaxation time - exceeds
the available experimental time. The understanding of the freezing
phenomenon and of the low temperature - or glassy - phase has been
boosted by the
introduction and the resolution of models such as the
spherical p-spin model, or the mode-coupling theory~\cite{Got89}. In
these models, the dynamics stops at some critical temperature $T_c$,
the idealised
glass transition temperature. The dynamics of the spherical p-spin
model has been solved in the limit of very large waiting
times~\cite{Cug93}, which
lead to the following phenomenology in the glassy
phase. In this phase a sharp separation of timescales occurs, with
two distinct time sectors~\cite{Cug02}, which in turn correspond to fast
and slow degrees of freedom. The former correspond to local (in time)
equilibration of local (in space) degrees of freedom and is often 
pictured by a rattling inside ``cages'', which respond to a small
external perturbating field through Fluctuation Dissipation Theorem
(FDT). The latter rule the slow structural relaxation, following a slow
descent in the energy landscape~\cite{Kur96}, and their response 
can be described using a {\it modified} FDT, in which the temperature
$T$ is replaced by an {\it effective} temperature $T^*\geq T$.
Still,
old samples exhibit stationarity on finite time differences.
Another
striking consequence of the analytic solution of the p-spin model in
the limit of an infinite waiting time is that for very old samples
the correlation
function $C(t,t')$ (we will always choose $t>t'$, with $t$ written
on the left) takes the form ${\cal C}\left(\frac{h(t')}{h(t)}\right)$
when $t\rightarrow\infty$ with $t\sim t'$. This form has been used to
fit data in experiments and simulations of aging systems.
In many cases, $h(t)$ is fairly well approximated by a power law (``simple
aging'')~\cite{Vin96,Clo00}~; however better fits have often been 
obtained by using
$h(t)=\exp\left(t^{1-\mu}/(1-\mu)\right)$, with $\mu<1$ but
$\mu$ close to $1$ (``sub aging'')~\cite{Str78,Vin96,Bou97}.
The interpretation was that the times used for measurements were two
small and that longer waiting times would provide better
agreement with simple aging. However, the study of the p-spin (or related)
models has not helped resolving this issue, as in this case $h(t)$ is
left unknown by the dynamic equations in the aging regime. In other
words, in the aging regime, the dynamics is
invariant under strictly increasing time reparametrizations~\cite{Cug93,Ken01}.

In this Letter, we revisit the analytic solution
of~\cite{Cug93}, by unveiling a new time scale, which as we shall
show is crucial in order to understand the crossover from the stationary
to the aging regime, and which we shall call the ``scaling
regime''. It corresponds to times of the order $t_w^\beta$, where $t_w$
is the waiting time and $\beta$ a new exponent, and was studied in
detail in the spherical Sherrington-Kirkpatrick (SK) model in~\cite{Zip00}. 
Our main results are
the calculation of the aging function $h(t)$ and the description
of the large but finite
waiting time behaviour of the dynamic correlators. 

\section{The model} 
We focus on the celebrated spherical p-spin model is defined by the
Hamiltonian $H_p=\sum_{1\leq i_1\cdots i_p\leq N} J_{i_1\cdots
  i_p}\phi_{i_1}\cdots \phi_{i_p}$, where $N$ is the number of soft spins,
which verify the spherical constraint $\sum_i\phi_i^2=N$ and the $J_{i_1\cdots
  i_p}$'s are Gaussian random coupling with vanishing average and
variance $N^{p-1}/p!$. Except when specified, we shall consider only $p>2$, where a
discontinuous ideal glass transition occurs~\cite{Cri92}. The dynamic
equations for these 
models in the large $N$ limit involve the correlator
$C(t,t')=\langle\phi_i(t)\phi_i(t')\rangle$ and the
response $R(t,t')=\langle\frac{\delta\phi_i(t)}{\delta \eta_i(t')}\rangle$
to an external field $\eta_i$. It is convenient to use the
integrated response $F(t,t')=\int_{t'}^t ds\,R(t,s)$, which leads to
the following equations~\cite{Cug02}:
\begin{eqnarray}
\left(\partial_t+z(t)\right)F(t,t')&=&1+\int_{t'}^t ds\,\Sigma(t,s)
F(s,t'),\label{eqn:dyn1}\\
\left(\partial_t+z(t)\right)C(t,t')&=&\int_0^t ds\,\Sigma(t,s)
 C(t',s)+\int_0^{t'} ds\, D(t,s) \partial_sF(t',s),\label{eqn:dyn2}
\end{eqnarray}
where $\Sigma(t,s)=-\frac{p(p-1)}{2}C(t,s)^{p-2} \partial_sF(t,s)$,
$D(t,s)=\frac{p}{2}C(t,s)^{p-1}$ and the Lagrange multiplier
$z(t)=T+\frac{p^2}{2}\int_0^t ds\,C(t,s)^{p-1} \partial_sF(t,s)$ 
enforces the constraint $\sum_i \phi_i(t)^2=N$. There is no
characteristic timescale in Eqs. (\ref{eqn:dyn1},\ref{eqn:dyn2}), as
the time is expressed in units of the microscopic timescale $t_0$.

\section{The different time regimes} 
In the limit of large waiting time
$t'\rightarrow\infty$, the dynamics can be partly solved assuming the
sharp separation between the stationary and aging
regimes~\cite{Cug02}. In this
limit, when $t-t'\sim 1$, the solution of (\ref{eqn:dyn1}) and (\ref{eqn:dyn2}) is stationary:
$C(t,t')=C_{ST}(t-t')$ and $R(t,t')=R_{ST}(t-t')$, where $C_{ST}$ and
$R_{ST}$ are related by FDT: $T R_{ST}(\tau)=-dC_{ST}(\tau)/d\tau$ ($\tau>0$).
The stationary part of the correlation behaves at large $\tau$ as
$C_{ST}(\tau)=
\q+c_{ST}^{(1)}\tau^{-a}+c_{ST}^{(2)}\tau^{-2a}+\cdots$, where $a$
verifies $\frac{\Gamma(1-a)^2}{\Gamma(1-2a)}=T/(2T^*)$. On the
other hand, when $t-t'\sim t'$, the solution has the
form~\cite{Cug93} $C(t,t')={\cal C}\left(
\frac{h(t')}{h(t)}\right)$, $F(t,t')={\cal
  F}\left(\frac{h(t')}{h(t)}\right)$, which satisfy ${\cal
  F}(\lambda)=\frac{1-{\cal C}(\lambda)}{T^*}$. The effective
temperature $T^*$ verifies $T/T^*=\frac{(p-2)(1-q_{EA})}{q_{EA}}$,
while the Edwards-Anderson parameter satisfies
$\frac{p(p-1)}{2}\q^{p-2}(1-\q)^2=T^2$. Near $\lambda=1$:
${\cal C}(\lambda)=\q-c_{AG}^{(1)}(1-\lambda)^b
-c_{AG}^{(2)}(1-\lambda)^{2b}+\cdots$, where $b=1$ for the spherical
p$(>2)$-spin 
model (we keep $b$ when possible, as some of the results apply to a
wider class of models)~\cite{Cug96}.

Now we add an extra timescale to the picture. We assume the existence of
an exponent $\beta$ which governs the size of the plateau with $t'$,
or equivalently $t$. We describe time differences of order
$t^\beta$, interpolating between the stationary and aging regimes,
thus $\beta<1$. The existence of this scaling regime is confirmed by
numerics~\cite{Kim01} and by the analytic solution in the case
$p=2$~\cite{Zip00}. The thickness of the plateau is described by another
exponent $\alpha$ such that
$C(t,t')=\q+t^{-\alpha} g_1(\frac{t-t'}{t^\beta})+t^{-2\alpha}
g_2(\frac{t-t'}{t^\beta})+\cdots$ and $F(t,t')=\frac{1-\q}{T}+t^{-\alpha}
w_1(\frac{t-t'}{t^\beta})+t^{-2\alpha}
w_2(\frac{t-t'}{t^{\beta}})+\cdots$. This implies
$R(t,t')=t^{-\gamma}r_1(\frac{t-t'}{t^\beta})+t^{-\gamma-\alpha}r_2(\frac{t-t'}{t^\beta})
+\cdots$, with $\gamma=\alpha+\beta$, and $r_i(x)=\partial w_i(x)/\partial
x$. In addition, one also introduces $\tilde T(x)=-g_1'(x)/w_1'(x)$,
which goes from $T$ at small $x$ to $T^*$ at large $x$~; the
finiteness of $\tilde T(x)$ explains the occurrence of the same
$\alpha$ for $C$ and $F$.

\section{Matching the different regimes} The scaling regime interpolates
between the stationary regime at small arguments and the aging
regime at large arguments. This gives relations between different
exponents and scaling laws. First we take $\tau=t-t'=\epsilon
t^\beta$, with $t\rightarrow\infty$ and $\epsilon\ll 1$ (in this
order). Matching with the stationary regime gives
$t^{-\alpha}g_i(\epsilon)\sim_{\epsilon\rightarrow 0}
c_{ST}^{(i)}\epsilon^{-ia}t^{i\beta a}$, and thus
$g_i(\epsilon)\sim c_{ST}^{(i)}\epsilon^{-ia}$, $\alpha=\beta a$. It also gives
$w_i(\epsilon)\sim -c_{ST}^{(i)}/T \epsilon^{-ia}$. This is similar to
the analysis 
around the plateau in the high temperature phase~\cite{Got89}.
Second, we take $\tau=x t^\beta$
with $x\gg 1$ and match with the aging regime. On one hand $C(t,t')={\cal
  C}\left(\frac{h(t')}{h(t)}\right)= \q-c_{AG}^{(1)} (\tau
\phi(t))^b-\cdots$, with $\phi(t)=\frac{d\ln
  h(t)}{dt}$. On the other hand
$C(t,t')=\q+t^{-\alpha}g_1(x)+\cdots$. Comparing the dependences in $\tau$
and $t$ gives $g_i(x)\sim -c_{AG}^{(i)} (A_ix)^{bi}$, $w_i(x)\sim 
c_{AG}^{(i)} (A_ix)^{bi}/T^*$ and $\phi(t)\sim t^{-\mu}/A_i$, with
$\mu=\beta+\alpha/b$. On can conclude that $h(t)\sim
\exp\left(\frac{t^{1-\mu}}{A_1(1-\mu)}\right)$ , provided
$\phi(t)-t^{-\mu}\ll 1/t$. The corrections to $t^{-\mu}$ are given by
the corrections to the leading order of the $w_i$'s ($i\geq
2$)~\cite{And06} and can be neglected when $\mu+\alpha>1$. There is
numerical evidence (see below) that this is the case for p-spin models, 
and thus $\phi(t)=t^{-\mu}/A_1+o(t^{-1})$. This gives $h(t)\sim\exp\left(
\frac{t^{1-\mu}}{A_1(1-\mu)}\right)$, and thus
$0\leq\mu\leq 1$, as $h(t)$ must map infinity onto infinity. We set
$A_1$ to $1$, as changing $A_1$ to $A_1'$ is equivalent to
changing ${\cal C}(\lambda)$ to ${\cal C}(\lambda^{A_1/A_1'})$. 
This freedom to choose $A_1$ reflects that the miscroscopic timescale
is forgotten in the aging regime. It provides further support for the
absence of non-vanishing (except logarithmic) corrections
to $\ln h(t)\sim t^{1-\mu}/(1-\mu)$. Indeed, $\ln
h(t)=t^{1-\mu}/(1-\mu)+A_2 t^\delta+\cdots$, with $\delta>0$ would
fix the microscopic timescale $t_0$ through $A_2\propto t_0^{-\delta}$. 
This argument can even be used to give directly the form of
$h(t)$. Indeed, assuming ${\cal C}(\lambda)=\lambda^{\nu(t_0)}$,
$h(t)=\exp f(t/t_0)$ and $\partial_{t_0} C(t,t')=0$ in the aging
regime gives $f(u)=A(t_0)+B(t_0) u^c$~\cite{And06}. Thus $h$ is either a stretched
exponential or a power law, while other empirical forms~\cite{Bou97} are ruled out. Remark that in order
to obtain $h(t)$ by matching the different regimes, we have integrated
the large time behaviour, so
this in principle gives $h(t)$ only approximately. However
the meaning of the aging function is only asymptotic, and valid for
large $t$. Note that the relations we have derived between exponents
are very general, provided the expansions we used for the late
stationary and early aging regimes are valid.

\section{Equations for the scaling regime} Now we sketch the -
lengthy - derivation
of the equations for the scaling functions at the plateau, which will
be detailed elsewhere~\cite{And06}.  The following procedure gives equations for $g_1$
and $w_1$, but the exponents $\alpha$ and $\beta$ go away, and
$\beta$ remains unknown. We set $t-t'=x t^\beta$
in (\ref{eqn:dyn1}) and (\ref{eqn:dyn2}) and separate the integrals in
different time sectors. 
Then, the identification of different orders gives the
equations. In the mean time, one gets for large times
$z(t)=z(\infty)-z_1 t^{-\alpha}-z_2 t^{-2\alpha}-\cdots$,
where $z_1=0$ and $z_2>0$~\cite{And06}. The multiplier $z(t)$ being a one-time
quantity, its value is rather easy to get in numerics, and thus
provides a useful estimate for $\alpha$~; in addition, it also gives
the behaviour of the energy density ${\cal E}(t)=(T-z(t))/p$. Now we write
$\int_0^{t'}ds=\int_0^{\eta t}+\int_{\eta t}^{t'-\Lambda t^\beta}+\int_{t'-\Lambda
  t^\beta}^{t'-\epsilon t^\beta}+\int_{t'-\epsilon t^\beta}^{t'}$,
which separates $t-s$ in several distinct regimes:
very short waiting times, aging, scaling and stationary. Next we
take the limit $t\rightarrow\infty$, followed by $\eta\rightarrow 0$,
$\Lambda\rightarrow\infty$ and $\epsilon\rightarrow 0$. We then
identify all orders in $t^{-\alpha}$: taking the above limits leads to
divergences which cancel each other giving {\it en passant} the values
of the coefficients of the leading orders of the $g_i$'s
and $w_i$'s ($i\geq 2)$. We also write
$\int_{t'}^t=\int_{t'}^{t'+\epsilon_1 t^\beta}+\int_{t'+\epsilon_1
  t^\beta}^{t-\epsilon_2 t^\beta}+ \int_{t-\epsilon_2 t^\beta}^t$ and
send $\epsilon_1$ and $\epsilon_2$ to zero.
We now give the results of the identification of different
orders. First, the orders $t^0$ and $t^{-\alpha}$ of
(\ref{eqn:dyn1},\ref{eqn:dyn2}) give:
$z(\infty)=\frac{p(p-1)}{2} \q^{p-2}(1-\q)+\frac{1-\q^{p-1}}{T}$, and
we get at order $t^{-2\alpha}$:
\begin{eqnarray}
w(x)^2+\int_{x_0}^x dy\,w'(y)
\frac{g(y)}{T^*}+\int_0^x dy\,w'(y)\left(w(x-y)-w(x)\right)=C_F,
\label{eqn:scal1}\\
\int_{x_0}^\infty dy\,\left[(p-1)g(y)\left(w'(x+y)-w'(y)\right)+w'(y)\left(g(x+y)-g(y)\right)\right] \nonumber
\\\label{eqn:scal2}
+ (p-2)\int_0^\infty dy\,w'(x+y)\left(g(x+y)-g(y)\right) + \int_{x_0}^x dy\, w'(y)(g(x-y)+(p-2)g(y)) \\
+ \int_0^{x_0}dy\left[(p-1)g(y)w'(x+y)+w'(y)\left(g(x+y)+g(x-y)-2 g(x)\right)\right] \nonumber\\
+ \frac{g(x)^2}{2T^*}+2w(x_0)g(x) - \frac{(p-1)\nu}{T^*}q_{EA}g_\infty x +
\frac{(p-2)q_{EA}^2\nu^2}{2T^*}x^2=C_C,\nonumber 
\end{eqnarray}
where $x_0$ is such that $g(x_0)=0$, and $g_\infty$, $C_F$ and $C_C$ 
will be given somewhere else~\cite{And06} ($\nu$ is given below).
In order to obtain consistent equations, we had to make the assumption
that $C(t,t')$ vanishes faster than $t^{-2\alpha}$ when $t'$ is finite
and $t\rightarrow\infty$. This assumption is consistent with
the stretched exponential behaviour of $h(t)$, and has been checked numerically.
We thus ended up with coupled equations for $g_1$ and $w_1$. It
can be easily checked that $(\ref{eqn:scal1})$ is compatible with the
asymptotic behaviours obtained from the matching with the other
regimes~; the same check for $(\ref{eqn:scal2})$ is harder. 
In addition, integrating these equations from small
to large $x$ in principle provides the value of $c_{AG}^{(1)}$.
The p-spin model is a special case
where the form of the aging scaling function is known~; indeed
one has simply~\cite{Cug93}: ${\cal C}(\lambda)=\q \lambda^\nu$, which
gives $c_{AG}^{(1)}=-\nu$ (recall $A=1$). Thus, ${\cal C}$ is in
principle determined by $g_1$ and $w_1$, although we have not yet been
able to solve (\ref{eqn:scal1},\ref{eqn:scal2}) analytically or numerically.

\section{The case $p=2$} For $p=2$, $C(t,t')$ and $R(t,t')$ can be computed
exactly~\cite{Dea95}, and $T^*=\infty$. In addition the exponents of
the scaling regime are~\cite{Zip00} $\beta=4/5$ and $\alpha=2/5$. Moreover
the analysis must be slightly refined~\cite{And06}, as $g_1=0$ and correspondly
$c_{AG}^{(1)}=0$~; it gives $\mu=1$, which corresponds
to simple aging, as confirmed by the exact solution. Remark that here
one has $z_2=0$, while $z_3=-3/8$.

\begin{figure}[t]
\begin{center}
\includegraphics[width=.45\textwidth,clip]{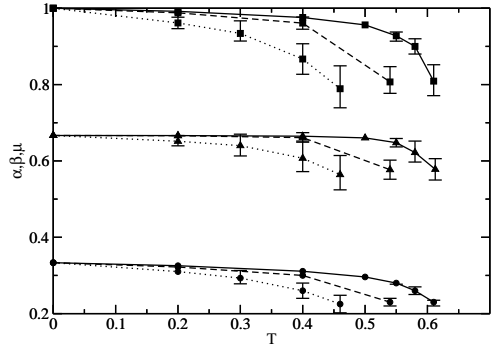}
\end{center}
\caption{\label{fig:exp} Exponents $\alpha$ (circles), $\beta$
  (triangles) and $\mu$ (squares)
  versus temperature for $p=3$ (solid lines), $p=4$ (dashed lines) and
  $p=10$ (dotted lines).}
\end{figure}

\section{Numerical check} 
\begin{figure}[t]
\begin{center}
\includegraphics[width=.45\textwidth,clip]{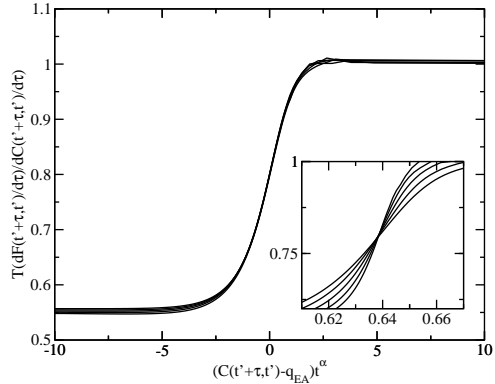}
\end{center}
\caption{\label{fig:scaling} Test of the scaling hypothesis for
$p=3$
  and $T=0.5$.  Inset: same as main, but with $C(t'+\tau,t')$ as
  horizontal coordinate~; the crossing occurs at $q$ and corresponds
  to $\tau=x_0 t^\beta$, with $g(x_0)=0$. Waiting times are $t'=2197$,
  $t'=4295$, $t'=8590$, $t'=17180$ and $t'=34359$.}
\end{figure}

As one exponent is still unknown, we have
checked the relations between different exponents using a code for numerical
integration of the equations provided by the authors
of~\cite{Bir06}. We start with
$p=3$. At $T=0.5$, we obtained $\alpha\approx 0.3$ from the fit
of $z(t)$, and $\beta\approx 0.66$ from the scaling of $t_\beta(t')$
which solves 
$C(t'+t_\beta(t'),t')=\q$. This is in very good agreement with $\alpha=\beta
a$ ($a\approx 0.448$) and gives $\mu\approx 0.96$, which provides a very
good collapse of the data in the aging regime, as found empirically
in~\cite{Kim01}.
The same test was carried out at several temperatures from $T=0$ to $T=0.61$
($T_c=\sqrt{3/8}\approx 0.6125$), providing very good agreement with
our analytic description. In Fig.~\ref{fig:exp}, the exponents
$\alpha$, $\beta$ and $\mu$
are shown (along with the same for $p=4$ and $p=10$). The observed
tendency is that they decrease with $p$ and $T$. In
addition, $\beta$ and $\mu$ saturate respectively to $2/3$ and $1$,
their values at $T=0$. The value $\mu(T=0)=1$ corresponds to approximate
aging slightly above $T=0$.
It is clear from Fig.~\ref{fig:exp} that
$\mu+\alpha>1$ is a reliable working assumption. Finally, we have checked
numerically the collapse of the data when using the scaling we
introduced for the plateau regime: in Fig~\ref{fig:scaling}, $-T\partial
F/\partial C$ is represented versus $(C(t'+\tau,t')-q_{EA}) t'^\alpha$
parameterised by $\tau$ for several waiting times, for $p=3$ and
$T=0.5$. A consequence of our scaling hypothesis is that a master
curve is obtained, identical to the curve of $T/\tilde T(x)$ versus
$g(x)$, which shows the onset of FDT violation during the scaling regime.

\section{Discussion} In many glassy systems, fits of the aging regime
were done using $h(t)\sim t$ (or any power law), giving reasonable
agreement. It was however observed that $h(t)\sim
\exp\left(t^{1-\mu}\right)$ gives better collapse of the
thermo-remanent magnetisation in spin-glass experiments, with
$\mu$ slightly below $1$~\cite{Str78,Vin96}. An
interpretation of this fact was that the waiting time was not large
enough for the aging scaling to be seen well. In the case of the
p-spin model, as we have shown, the stretched exponential form
is related to the power law dependance of the timescale of the
plateau with the waiting time and to the forgetting of the microscopic
timescale in the aging regime. In addition, we have found $\mu<1$ from
numerical estimates, which appears as an evidence against simple
aging. However 
$\mu$ departs significantly from $1$ in the vicinity of the
critical point, and the numerical estimation of the exponents may be
flawed by critical effects. For instance, the exponents shown on
Fig.~\ref{fig:exp} seem to have a cusp at $T_c$, which could be an
indication that the scaling regime is strongly influenced by the cusp
of $q_{EA}$ at $T_c$~\cite{Got89}, as the position of the plateau
itself fluctuates a lot at the approach to $T_c$.
A simple way to discriminate such
effects would be to compute the variation of $\beta$ with $T$. Indeed
our numerical results suggest that $\beta$ decreases with $T$, while
$\beta(T)$ obtained by imposing $\mu=1$ at all temperatures - which gives
$\beta(T)=1/(1+a(T))$ - is an increasing function. It remains that for
fixed $t'$, our data and those of~\cite{Kim01} are clearly compatible 
with stretched exponentials rather than power laws. The scenario  
$\mu<1$ is at least relevant for the time regimes numerically accessible.

\section{Conclusion} In this Letter, we have shown that the matching of
the stationary and aging regimes in glassy systems requires the
precise description of the intermediate scaling regime, which is
defined at large but finite waiting time. We have shown that the
natural assumption that in this regime the correlators obey power law
scaling leads to a stretched exponential form for the aging
function. In addition, the exponent is related to the one giving the
duration of the plateau. The assumptions and ans\"atze which have been
made in order to get information about the scaling regime have been
checked with very good accuracy using numerical integration of the
exact dynamic equations. The consequences of these results are
threefold. First, it shows the p-spin model as an archetype system
where the aging
function is close to a power law, but different, at least in the
accessible time regimes. This provides
a motivation for revisiting experimental and numerical data fits where
the same phenomenon was observed. Second, our
calculation provides explicit corrections at large but finite times to
the solution given in~\cite{Cug93}. This may indicate the route to
follow in order to study the dynamics of systems where the ideal
transition does not really exist, and where an aging behaviour, with a
waiting time dependent effective temperature was found. Here a
calculation at large, but not infinite waiting time is clearly
necessary.
Third, we hope that our treatment may lead to a proper mathematical
approach to the solution of the dynamic equations. Indeed, it has
recently been proven rigorously~\cite{Ben05} that the dynamic
equations governing
the dynamics in the (thermodynamic) limit $N\rightarrow\infty$ derived
in~\cite{Cug93}. The important challenge is now to compute the
exponent $\beta$, i.e. the length of the plateau. Indeed, it is
important to confirm whether or not $\mu<1$ is not a numerical effect
due to critical fluctuations.  
Finally, our analysis remains to be adapted to models
with continuous replica symmetry breaking such as the SK model~\cite{She75}.

{\em Acknowledgements - } The code for numerical integration was kindly given
by K. Myazaki, whom we warmly thank for his technical support. We also
thank G. Biroli and J.-P. Bouchaud for useful and stimulating
discussions, as well as L. Berthier and J.-M. Luck for
useful comments on the manuscript.


\begin{thebibliography}{99}
\bibitem{Deb01}{P. G. Debenedetti and F. H. Stillinger, {\em Nature}
  {\bf 410}, 259 (2001).}
\bibitem{Kov63}{A. J. Kovacs, {\em Adv. Polym. Sci.} {\bf 3}, 394 (1963).}
\bibitem{Joh98}{K. Johnason et al., {\em Phys. Rev. Lett.} {\bf 81},
  3243 (1998).}
\bibitem{Hur95}{M. M. Hurley and P. Harrowell, {\em Phys. Rev. E} {\bf
    52}, 1694 (1995)~; C. Bennemann et al., {\em Nature} {\bf 399}, 246
    (1999), C. Donati et al., {\em J. Non-Cryst. Solids} {\bf
    307}, 215 (2002)~; S. Whitelam, L. Berthier and J. P. Garrahan, {\em
    Phys. Rev. Lett.} {\bf 92}, 185705 (2004)~; L. Berthier, {\em
    Phys. Rev. E} {\bf 69}, 020201 (2004)~; G. Biroli and
    J.-P. Bouchaud, {\em Europhys. Lett.} {\bf 67}, 21 (2004).}
\bibitem{Got89}{W. G\"otze, {\em in ``Liquids, freezing and the glass
      transition'', Proceedings of Les Houches summer school} (1989).}
\bibitem{Cug93}{L. F. Cugliandolo, and J. Kurchan, {\em
    Phys. Rev. Lett.} {\bf 71}, 173 (1993)~; L. F. Cugliandolo, and
    J. Kurchan, {\em Progress. in Theor. Phys.}{\bf 126}, 407 (1997).}
\bibitem{Cug02}{L. F. Cugliandolo, {\em in ``Slow relaxations in
      condensed matter'',  proceedings of Les Houches
    summer school} (2002).}
\bibitem{Kur96}{J. Kurchan and L. Laloux, {\em J. Phys. A} {\bf 29},
  1929 (1996).}
\bibitem{Vin96}{E. Vincent et al., {\em in ``Proceedings of the Stiges
    Conference on Glassy Systems''}, (1996), {\em cond-mat/9607224}.}
\bibitem{Clo00}{M. Cloitre, R. Borrega and L. Leibler, {\em
    Phys. Rev. Lett.} {\bf 85}, 4819 (2000)~; R. J. M. d'Arjuzon,
    W. Frith and J. Melrose, {\em Phys. Rev. E} {\bf 67}, 061404 (2003).}
\bibitem{Str78}{L. C. E. Struik, {\em Physical Aging in Amorphous
    Polymers and Other Materials}, Elsevier, Houston (1978)~; 
    M. Ocio, M. Alba and J. Hammann, {\em J. Physique
    Lett. (France)} {\bf 46}, L1101 (1985)~; M. Alba et al., {\em
    J. Appl. Phys.} {\bf 61}, 3683 (1987).}
\bibitem{Bou97}{J.-P. Bouchaud et al., {\em in "Recent progress in
      random magnets"}, A.P. Young ed., World Scientific (1997).}
\bibitem{Ken01}{M. P. Kennet and C. Chamon, {\em Phys. Rev. Lett.}
  {\bf 86}, 1622 (2001).} 
\bibitem{Cri92}{A. Crisanti and H.-J. Sommers, {\em Z. Physik B} {\bf
    87}, 341 (1992).}
\bibitem{Cug96}{L. F. Cugliandolo and P. Le Doussal, {\em
    Phys. Rev. E} {\bf 53}, 1525 (1996).} 
\bibitem{Kim01}{B. Kim and A. Latz, {\em Europhys. Lett.} {\bf 53},
  660 (2001).}
\bibitem{Zip00}{W. Zippold, R. K\"uhn and H. Horner, {\em
    Eur. Phys. J. B} {\bf 13}, 531 (2000).}
\bibitem{And06}{ A. Andreanov and A. Lef\`evre, unpublished.}
\bibitem{Dea95}{L. F. Cugliandolo and D. S. Dean, {\em J. Phys. A}
  {\bf 28}, 4213 (1995).}
\bibitem{Bir06}{G. Biroli et al., {\em cond-mat/0605733} (2006).}
\bibitem{Ben05}{G. Ben Arous, A. Dembo and A. Guionnet, {\em
    cond-mat/0409273} (2004).}
\bibitem{She75}{D. Sherrington and S. Kirkpatrick, {\em
    Phys. Rev. Lett.} {\bf 35}, 1792 (1975).}
\end{thebibliography}
\end{document}